\documentclass[aip,twocolumn,floatfix,11pt]{revtex4-1}
\usepackage{bm}
\usepackage{graphicx}
\usepackage{amsmath}
\usepackage{amssymb}
\usepackage{setspace}
\begin{document}

\title{Computational Studies of Multiple-particle Nonlinear Dynamics in a Spatio-Temporally
Periodic Potential}

\author{Owen D. Myers}
\affiliation{University of Vermont, Burlington, Vermont 05405, USA.}
\affiliation{Materials Science Program}
\author{Junru Wu}
\affiliation{Materials Science Program}
\affiliation{Department of Physics}
\affiliation{University of Vermont, Burlington, Vermont 05405, USA.}
\email{Junruwu@gmail.com.}
\author{Jeffrey S. Marshall}
\affiliation{School of Engineering}
\affiliation{University of Vermont, Burlington, Vermont 05405, USA.}
\author{Christopher M. Danforth}
\affiliation{Department of Mathematics and Statistics}
\affiliation{Vermont Complex Systems Center}
\affiliation{University of Vermont, Burlington, Vermont 05405, USA.}

\date{\today}

\begin{abstract}
The spatio-temporally periodic (STP) potential is interesting in Physics due to the intimate
coupling between its time and spatial components. In this paper we begin with a brief discussion of
the dynamical behaviors of a single particle in a STP potential and then examine the dynamics of
multiple particles interacting in a STP potential via the electric Coulomb potential. For the
multiple particles’ case, we focus on the occurrence of bifurcations when the amplitude of the STP
potential varies. It is found that the particle concentration of the system plays an important role;
the type of bifurcations that occur and the number of attractors present in the Poincar\'{e} sections
depend on whether the number of particles in the simulation is even or odd. In addition to the
nonlinear dynamical approach we also discuss dependence of the squared fractional deviation of
particles’ kinetic energy of the multiple particle system on the amplitude of the STP potential
which can be used to elucidate certain transitions of states; this approach is simple and useful
particularly for experimental studies of complicated interacting systems.  

\end{abstract}


\maketitle 

\section{Introduction}
Studies of nonlinear periodically driven systems are important to understanding the fundamental
physics of many useful and interesting phenomena in applications of lasers \cite{nonlin_laser},
driven ratchets \cite{ratchet}, hydrophilic particles on the surface of water waves
\cite{faraday_surf_1,faraday_surf_2,faraday_surf_3}, Josephson junctions \cite{josephson_prl}, etc.
If bodies moving in periodically-driven systems are allowed to interact, they can display a wealth
of interesting physical phenomena associated with complex systems
\cite{cell_aut_1,cell_aut_2,chimera}. Studies of the interaction among oscillators, bodies, nodes,
etc., in periodic systems, and of how bodies in such systems collectively react to environmental
forcing, can be valuable in a variety of fields from neuroscience\cite{neuro} to driven Josephson junction
arrays \cite{josephson_arr}.  

Interest in spatio-temporally periodic
potentials (STP) began in 1951, when Kapitza published two papers \cite{kapitza_1,kapitza_2} on a
planar pendulum with an oscillating suspension point, which is often referred to as the
parametrically driven pendulum, or simply Kapitza's pendulum. The most interesting feature of this
simple system is that under certain conditions, the pendulum stands stably in the inverted position.
The change in stability of the inverted position through the oscillation of the suspension point is
an example of what is known as dynamic stabilization, in which an inherently unstable system can be
stabilized by periodic forcing.  

The motion of a single particle immersed in a STP potential may be solved analytically using a
simple approximation. For instance, for small $x$ we may approximate $f(\vec{x})$ with a harmonic
oscillator potential resulting in equations of motion that can then be solved using Floquet theory
\cite{lefschetz}.  Linearization of the equation of motion in the limit of zero dissipation for
one-dimensional motion allows the equation of motion to be expressed as the Hill equation,
$\ddot{x}+g(t)x=0$. If $g(t)=\cos(\omega t)$, this equation reduces to the Mathieu equation
\cite{mclachlan}. Both the Hill and Mathieu equations have been studied extensively due to the
interesting properties that they display and due to the many applications that can be associated
with these equations, including the quantum pendulum \cite{qm_pendulum}, ion traps
\cite{ion_trap_rev}, and oscillations of a floating mass in a liquid \cite{mathieuapp}.  When the
oscillations of $g(t)$ are fast compared to the natural
frequency of oscillations when $g(t)$ is set to its maximum value, the method of averaging can be
used \cite{kry_bogo_avg, MB_E_loc}.

The interesting physical behavior of Kapita's stable inverted
pendulum has enticed several researchers to study the nonlinear case both experimentally
\cite{prl_pen_experiment, control_chains_kap_pen, exp_fuzzy_control_pen} and numerically
\cite{hopf_inverted, global_bif_pen, kiminvertedpen, on_vert_pen, subharm_res_pen,
chaotic_behaviour_par_pen, control_chains_kap_pen}. For systems composed of particles in a STP
potential, only a small number of publications have examined the nonlinear multiple particle
dynamics accounting for the multi-particle interactions. Some examples of papers treating this
subject include a study of the motion of hydrophobic/hydrophilic particles on the surface of Faraday
waves \cite{faraday_surf_2,faraday_surf_3}, multiple charged particles in an STP potential generated
by an electric curtain \cite{masudastand,liu_jeff,incline_ec}, and multiple particles in a
periodically forced straining flow \cite{marshall2009}. These previous studies have considered very
large numbers of particles and they have focused on the overall particle motion. In the current
paper, we instead examine the dynamics of a relatively small number of particles in a STP potential
using a dynamical systems point of view. Specifically, we seek to relate the nonlinear systems
dynamics with multiple particles to the bifurcations and stability of single particles.

\section{Methods}
The current computational study examines a one-dimensional (1D) system with multiple particles
interacting through a repulsive electrostatic $1/r$ potential in an external STP potential field.
The STP potential is 

\begin{equation}
\Phi = -A\cos{x}\cos{t},
\label{eqn:1Dpotential}
\end{equation}

\noindent which produces equations of motion analogous to the parametrically driven pendulum in the
horizontal plane. The coefficient $A$ is the potential amplitude, and the distance coordinate $x$
and time coordinate $t$ are non-dimensionalized using the wavenumber $k$ and the STP driving
frequency $\omega$, respectively.  

The driving force ($F_{\Phi}=-\nabla \Phi$) has the form of a standing wave with oscillation
amplitude $A$. The dimensionless wavelength $\lambda$ and the oscillation period $T$ are both
equal to $2\pi$. The system is assumed to be periodic over $n \lambda$, where $n$ is an integer, so
we can define the concentration $\sigma$ as $N/n$ where $N$ is the number of particles in the
simulation. For simplicity, all particles are assumed to carry the same charge and mass, where the
non-dimensionalization is performed such that the dimensionless mass is equal to unity. Damping is
proportional to particle velocity with a dimensionless damping coefficient $\beta$. In the current
paper, we focus on the effect of the parameter $A$, and therefore maintain constant values of the
other dimensionless parameters - the damping parameter and the dimensionless particle charge. These
latter two parameters are set equal to $\beta=0.6$ and $q=1$ throughout the paper.  These values for
$\beta$ and $q$ are chosen because they are realistic for systems similar to those discussed in
\cite{masudastand,liu_jeff,incline_ec}, after being dimensionalized.

The force on a particle located at $x_i$ imposed by a particle located at $x_j$, denoted by
$F_{ij}$, is calculated with periodic boundary conditions. To address the forces imposed by long
range interactions, we consider an infinite sequence of image systems using Ewald summation method
\cite{long_range}, giving 

\begin{equation}
    \label{eqn:particle_force_sum}
    F_{ij}=\frac{q^2 r_{ij}}{\|r_{ij}\|^3}+q^2\sum_{\nu=0}^\infty \frac{1}{(2\pi \nu 
    - r_{ij})^2} - \frac{1}{(2\pi \nu + r_{ij})^2} ,
\end{equation}

\noindent where $r_{ij}=x_i - x_j$. The sum in (\ref{eqn:particle_force_sum}) is convergent and may
be written as a polygamma function $\psi^m(z)$ with series expansion

\begin{equation}
    \label{eqn:polygamma}
    \psi^{(m)}(z)=(-1)^{m+1}m!\sum_{\nu=0}^{\infty}\frac{1}{(z+\nu)^{m+1}}.
\end{equation}

\noindent Using (\ref{eqn:polygamma}), we can express $F_{ij}$ as

\begin{equation}
\begin{split}
    \label{}
    & F_{ij} = \frac{q^2 r_{ij}}{\|r_{ij}\|^3} - \left( \frac{q^2}{n\lambda} \right)^2 \\
    & \times \left( \psi^{(1)}(1+r_{ij}/\lambda)-\psi^{(1)}(1-r_{ij}/\lambda) \right)
\end{split}
\end{equation}

\begin{figure}[h]
    \centering
    \includegraphics[width=8.5 cm]{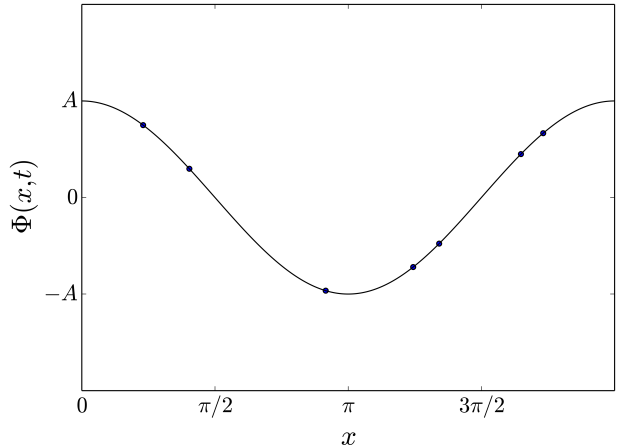}
    \caption{Example of the system with $N=7$ and $A=2.758$, depicted as particles on the surface of a standing wave (multimedia view). \label{fig:sup_still}}
\end{figure}

\noindent The equation of motion for the $i^{th}$ particle is given by

\begin{equation}
    \ddot{x}_i =  - \beta \dot{x_i} + F_{\Phi} + \sum_{j \neq i}^{N}F_{ij}   
\label{eqn:motion}
\end{equation}

\noindent where the second term on the RHS is due to the imposed STP potential field and the third
term on the RHS is the particle interactions. An example of the system containing seven particles may be
found in Fig. \ref{fig:sup_still} (multimedia view).

\subsection{The Phase Space}
For $N$ particles in an autonomous system, the degrees of freedom (or dimension) of
the phase space is $2 d N$, where $d$ is the dimension of the physical system. In STP problems,
there is an explicit time dependence in the potential and therefore the system is non-autonomous.
Non-autonomous systems may be transformed into autonomous form by introducing an extra degree of
freedom, which in a first-order system is given by $x_3 = t$. Though this seems to be a trivial
representation of time, this autonomous formulation is necessary when distinguishing types of
bifurcations. This augmented system thus has $2 d N + 1$ degrees of freedom, which constitutes the
``full phase space''.

\subsection{Poincar\'{e} Sections}
The standard choice for making Poincar\'{e} sections in driven systems is a time map taken at the
system driving period.  Time maps are stroboscopic views of a trajectory expressed as $x(t=2\pi n)$
when $n$ is a positive integer. A Poincar\'{e} section includes any point where a
continuous trajectory transversely intersects a subspace of the full phase space
\cite{guckenheimer}. Time maps, as defined above, will produce Poincar\'{e} sections with dimension
one less than the dimension of the full phase space. In a time map, the path of a particle is always
transverse to the $x-\dot{x}$ plane, and therefore a point of intersection of the trajectory with
this plane is a convenient sub-space that satisfies the criterion necessary to be a Poincar\'{e}
section.   

\subsection{Kinetic Energy Fluctuations}
It is well known that as a system approaches a bifurcation point, it may take longer for transients
of the relevant quantity to die out or for the system to recover from an external perturbation
\cite{crit_trans}. This behavior is known as the critical slowing down phenomena. Most real systems are
subject to some natural perturbations, and these perturbations can become particularly apparent near
the bifurcation points. Measuring the increase of the variance in a physical quantity can therefore be
used as a method to predict the presence of a bifurcation point \cite{crit_trans}. The model
system considered in the current paper has no external perturbations, aside from computer round-off
error. In the limit $t\rightarrow \infty$, the damped system would be expected to settle into an
attractor, but the finite time of real simulations ensures the presence of small fluctuations in
``residual'' transients. In other words, multiple particle systems have a ``large'' number of degrees of
freedom, therefore some small trace of the initial transient behavior (residual transients) will
most likely be detectable. The amount of residual transients may be found in the kinetic energy
fluctuations. It is known that the kinetic energy fluctuations may contain some information about
the "effective number of degrees of freedom" \cite{gran_gas}. The more degrees of freedom, the more
residual transients will be present. This correspondence between the effective degrees of freedom
and the kinetic energy fluctuations is what makes the kinetic energy fluctuations an interesting
quantity to examine. 

The square of the deviation of the particle kinetic energy is given by $(\Delta KE)^2
\equiv \langle KE^2 \rangle - \langle KE \rangle ^2$,

\begin{equation}
    (\Delta KE)^2 = \frac{1}{4} \sum_{i,j}^N (\langle v_i^2 v_j^2 \rangle - \langle v_i^2 \rangle
    \langle v_j^2 \rangle),
\label{eqn:varience}
\end{equation}

\noindent and $v_i$, $v_j$ denote the $i^{th}$ and $j^{th}$ particle velocities, respectively. The
average is calculated as $\langle KE \rangle^2 = \frac{1}{4}\sum_{i,j}^N \langle v_i^2 \rangle
\langle v_j^2 \rangle$. The normalized squared deviation of the kinetic energy is given by 

$$\delta_{KE}\equiv \frac{\Delta KE}{\langle KE \rangle^2}.$$

\section{Results}
\subsection{Single Particle Overview}
\begin{figure}[htp]
    \centering
    \includegraphics[width=8.5 cm]{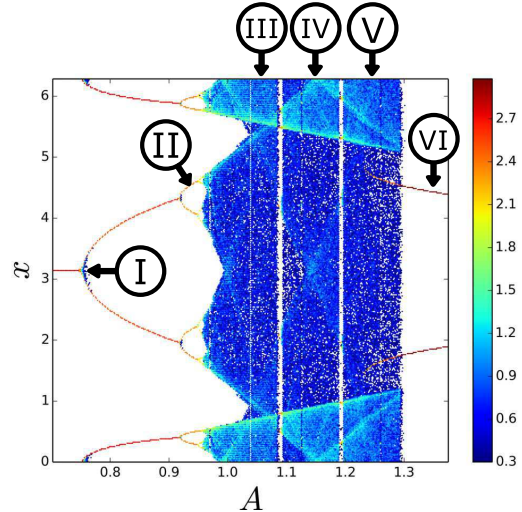}
    \caption{Bifurcation diagram formed by taking a two-dimensional histogram ($300\times 300$ bins) of the
    final Poincar\'{e} section of 1830 trajectories with different initial conditions for 300 different
    values of $A$.  The gray scale (color online) represents the base 10 logarithm of the number of
    particles in a bin. The Roman numerals are listed here for comparison with $\delta_{KE}$ shown
    in Fig. \ref{fig:N1_fluctuations}.
    \label{fig:bif_sin_1D}}  
\end{figure}

The dynamics of a single particle immersed in the one-dimensional STP potential $\Phi$, given by
(1), are similar to the dynamics of the parametric pendulum. In this paper, we only discuss dynamics
for the first bifurcation sequence leading to the chaotic regime, even though there are many consecutive
regimes of stable limit cycles bifurcating into chaotic trajectories. In Fig. \ref{fig:bif_sin_1D},
the first bifurcation sequence is shown for an ensemble of initial conditions. Table \ref{table:bif}
lists the type of bifurcations, the critical values of $A$ at which each bifurcation occurs, and the
period of the limit cycle following each bifurcation. The table is truncated after the $6^{th}$
bifurcation due to numerical resolution limitations for distinguishing bifurcation onset in a small
volume of the phase space. For $0<A<A_{c1}$, a particle will move toward and equilibrate at the
antinodes of the potential $\Phi(x,t)$ (i.e., the maxima of $\cos{x}$). As $A$ is increased, the
fixed points in the Poincar\'{e} section bifurcate in a supercritical flip bifurcation leading to a
period-2 limit cycle for $A_{c2}<A<A_{c3}$. This transition is not a Hopf bifurcation because the
explicit time dependence in the equations of motion must be considered as a degree of freedom to the
phase space. Consequently, what might appear as a fixed point in the $x-\dot{x}$ phase space in a
bifurcation diagram is actually a period-1 trajectory in the full phase space. We prove this using
Floquet theory, by numerically calculating the stability multipliers. Both of the two non-trivial
stability multipliers have no imaginary component close to the bifurcation point. At the
bifurcation, one stability multiplier becomes smaller than -1 while the other remains close to zero,
indicating a period-doubling supercritical flip bifurcation. This stability multiplier passing
through -1 is shown in Fig. \ref{fig:N1_fluctuations}a, where it is denoted with a Roman numeral I.
The values of $A$ for which the first six bifurcations occur, shown in Table \ref{table:bif},
indicate a period-doubling cascade route to chaos. The computed values yield a Feigenbaum constant
of 4.00 with an upper error bound of 6.00 and a lower bound of 2.89. The accepted value of 4.669 for
period-doubling bifurcations \cite{thompson} is within the error bounds. The Feigenbaum constant
$\mathcal{F}$ is evaluated with

\begin{equation}
	\mathcal{F}= \lim_{n\rightarrow \infty} \frac{A_{cn-1}-A_{cn-2}}{A_{cn}-A_{cn-1}}
\end{equation}

\noindent where $A_{cn}$ is the $n^{th}$ critical value of $A$ for which a period-doubling
bifurcation occurs. The two lines coming out of the chaotic region in Fig.  \ref{fig:bif_sin_1D} are
each attractors representing stable propagating trajectories, one with a positive velocity and one
with a negative velocity. These propagating trajectories travel across $\lambda$ once per period of
the driving potential field. 

\begin{table}[htp]
    \caption{Bifurcations}
    \centering
    \begin{tabular}{c c c c}
    \hline\hline
    $A_{cn}$ &Bifurcation & $A\pm 5e-5$ & New Period \\
    \hline
    $A_{c1}$ &Supercritical Flip& $ 0.75365 $ & $2$ \\
    $A_{c2}$ &Cyclic Fold       & $ 0.91875 $ & $2$ \\
    $A_{c3}$ &Supercritical Flip& $ 0.94985 $ & $4$ \\
    $A_{c4}$ &Supercritical Flip& $ 0.95650 $ & $8$ \\
    $A_{c5}$ &Supercritical Flip& $ 0.95790 $ & $16$ \\
    $A_{c6}$ &Supercritical Flip& $ 0.95825 $ & $32$ \\ [1ex]
    \hline
    \end{tabular}
    \label{table:bif}
\end{table}

\subsection{Kinetic Energy Fluctuations of One Particle}
\begin{figure}[htp]
    \centering
    \includegraphics[width=8.5cm]{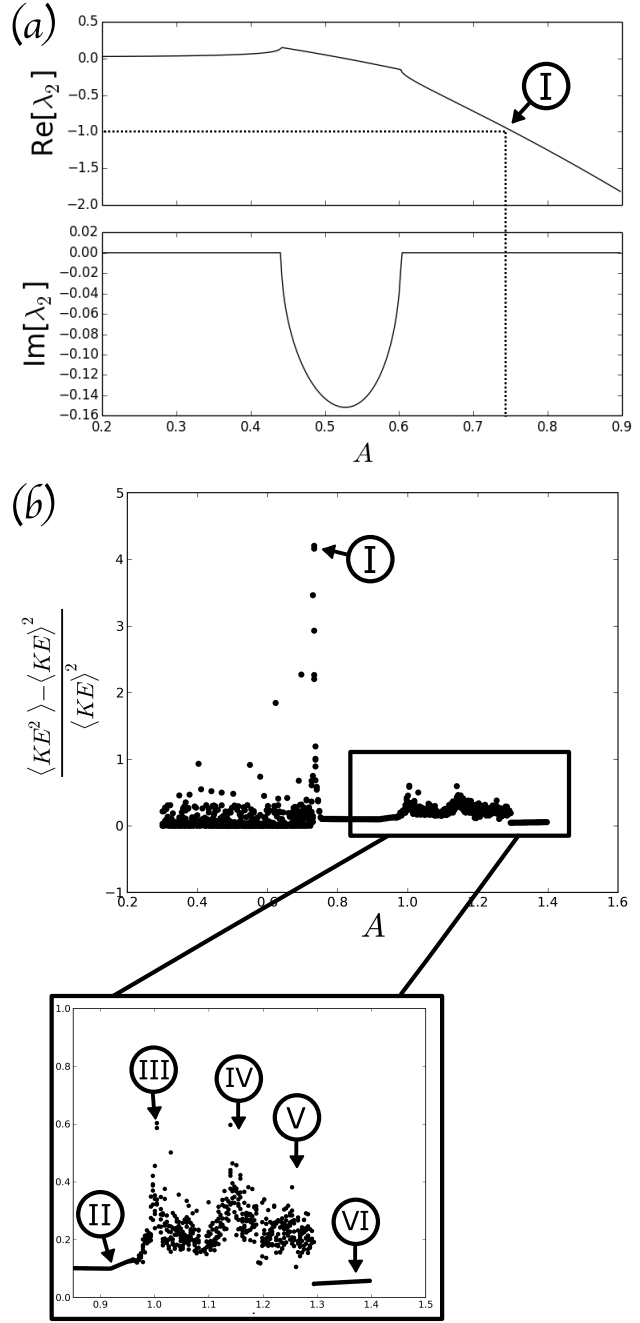}
    \caption{Single particle: (a) One of the two Floquet stability multipliers for the $x=\pi$ fixed
point as a function of the potential oscillation amplitude as it is increased through the first
bifurcation point. (b) Kinetic energy fluctuations. Roman numerals for comparison with the
bifurcation diagram in Fig. \ref{fig:bif_sin_1D} \label{fig:N1_fluctuations}} 
\end{figure}

Before going to the multi-particle case, it is informative to compare the bifurcation diagram (Fig.
\ref{fig:bif_sin_1D}) to the calculation of $\delta_{KE}$ for a single particle, which is shown in
Fig. \ref{fig:N1_fluctuations}. We also show a Floquet stability analysis of the fixed point at
$x=\pi$ through $A_{c1}$ for comparison. Floquet stability analysis is a powerful tool in analyzing
bifurcations, but it is not easily applied to multiple particle systems. It has been applied to
coupled Kapitza pendulums by \cite{coupled_invert_pen}. For a description of single particle
stability analyses, we refer the reader to \cite{kiminvertedpen} and \cite{mine}, which are both
studies of similar systems and use the Floquet technique to study bifurcations. In Fig.
\ref{fig:N1_fluctuations}a, the real and imaginary components of the stability multiplier that
causes the bifurcation (one of the two complex Floquet stability multipliers $\lambda_1$ and
$\lambda_2$) are plotted as $A$ is increased through $A_{c1}$. In Fig. \ref{fig:N1_fluctuations}b,
$\delta_{KE}$ is plotted as $A$ is increased through the full range shown in the bifurcation diagram
in Fig. \ref{fig:bif_sin_1D}. 

In Fig. \ref{fig:bif_sin_1D} and Fig. \ref{fig:N1_fluctuations}, the key regions associated with
different system behaviors have been identified using Roman numerals. For small values of $A$, Fig.
\ref{fig:N1_fluctuations}b shows a wide range of scattered points. However, the particle exhibits
very little motion within this range of small $A$ values. As $A$ is increased, there exists a peak
in the fluctuations near $A_{c1}$, which is a consequence of the critical slowing down phenomenon
(region I). As $A$ is increased past $A_{c1}$, the fluctuation amplitude is relatively constant
until $A$ approaches $A_{c2}$, where a kink is observed (region II). When $A$ is in the chaotic and
near-chaotic regimes (regions III,IV,V), the fluctuations increase in amplitude and are irregular,
as shown in the inset in the figure. The two regions where the fluctuation amplitude decreases
markedly in this inset correspond to the two periodic windows seen in Fig. \ref{fig:bif_sin_1D}. At
the end of the chaotic regime, there is a discontinuous jump in the fluctuations to a comparatively
small and relatively constant value (region VI). This last section shows the transition to
propagating trajectories, and we will see that this feature is present in all cases where this
transition occurs. Under closer inspection, region VI overlaps with region V because, just as in the
bifurcation diagram, the propagating trajectories exist simultaneously with the chaotic regime for a
small range of $A$.

\subsection{Integer Concentrations}
\begin{figure*}[htp]
    \centering
    \includegraphics[width=17 cm]{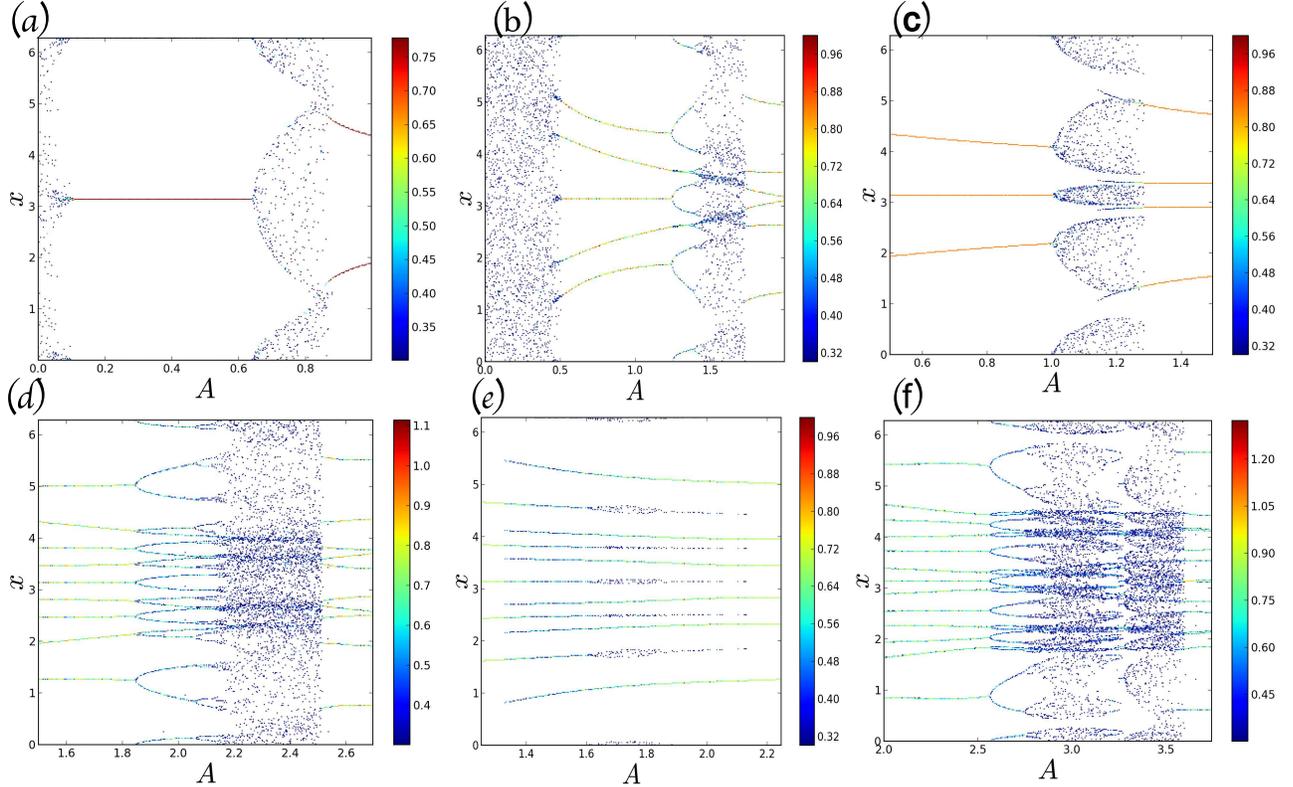}
    \caption{Multiple particle bifurcation diagrams made by: (1) initializing particles at random
    initial positions with $\dot{x}(0)=0$ , (2) projecting the Poincar\'{e} section onto the position
    axis after 150 cycles and plotting against the value of $A$, and (3) repeating the process for $400$
    values of $A$ in the range of interest. The number of particles used in each bifurcation diagram is
    (a) $N=2$, (b) $N=3$, (c) $N=4$, (d) $N=5$, (e) $N=6$, (f) $N=7$. Note that the range of A values
    explored differs in each figure. \label{fig:multi_bifs}} 
\end{figure*}

The bifurcation diagrams for multiple interacting particles, with $N=2,3,4,5,6,7$, are shown in Fig.
\ref{fig:multi_bifs}. The increased degrees of freedom that occur for $N>1$ make it difficult to
investigate an ensemble of initial conditions that exhaustively fill the phase space.  We use random
positions distributed with even probability across $x$ (with $\dot{x}(0)=0$) as initial conditions
for each run to explore a set of possible initial conditions. The bifurcation plots are made by
taking the last Poincar\'{e} section after 150 driving cycles of a simulation, projecting it onto
the position axis, and then plotting the positions against the value of $A$ used in that simulation.
For very small $A$, the final Poincar\'{e} sections are scattered because, for these values,
transients die out very slowly. For larger $A$, there are clearly defined points in the Poincar\'{e}
sections that denote limit cycles in the full phase space. For the rest of the paper, the stable
limit cycles in the Poincar\'{e} sections are referred to more generally as attractors.  At first
glance, Fig. \ref{fig:multi_bifs} appears to indicate a larger number of particles for odd values of
$N$ than it does for even values of $N$. As $A$ is further increased, a bifurcation occurs for all
the cases shown, although it is difficult to see in Fig. \ref{fig:multi_bifs}(e). For other values
of $A$, the diagrams in Fig. \ref{fig:multi_bifs} appear as scattered points, implying either
chaotic motion, high-period trajectories, or motion on a torus (after a Neimark bifurcation). Past
this scattered regime, the system collapses into a new stable regime that is
qualitatively similar to the propagating trajectories that occur after the chaotic regime in the
single particle case. 

\begin{figure}[htp]
    \centering
    \includegraphics[width=8.5 cm]{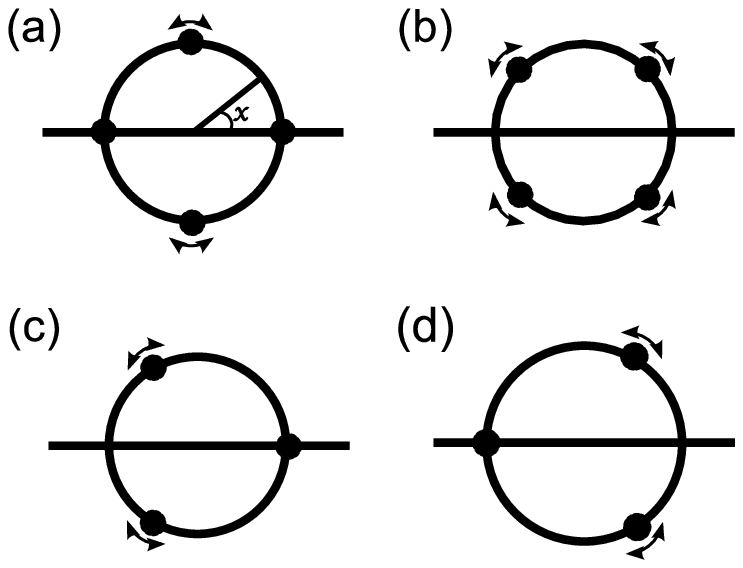}
    \caption{The position along the periodic domain is indicated by an angle
    around a circle. The black dots represent average particle positions before the first
bifurcation. The line bisecting the circle passes through the two antinodes of the potential field.
Part (a) shows the $N=4$ stable configuration which is an asymptotically stable fixed point in the
Poincar\'{e} sections of the full
phase space. Part (b) is an $N=4$ unstable configuration which is the unstable fixed point in the
Poincar\'{e} sections of the full phase space. Parts (c) and (d) show two different possible stable
configurations for $N=3$, both of which are stable fixed points in the Poincar\'{e} sections of the
full phase space.  \label{fig:even_odd}} 
\end{figure}

The discrepancy between the number of possible attractors for odd and even values of $N$ can be
explained by considering the relationship between the number of particles and the number of
antinodes in one period of the potential.  For the trajectory $(x_1(t),...,x_{2dN}(t))$, the number
of attractors that are observed for values of $A$ before the first bifurcation is equal to the
number of particles when $N$ is even and twice the number of particles when $N$ is odd. In other
words, there is one possible final state configuration of the whole system when $N$ is even and two
possible final state configurations of the system when $N$ is odd. To see why this occurs, we
examine in detail the cases of three and four particles per period. Figs. \ref{fig:even_odd}(a)(b) and
(c)(d) show cartoons of the final particle configurations for cases $N=4$ and $N=3$, respectively. In
these cartoons, the particle position in the periodic domain is drawn as an angle, so that motion of
the particle in the $x$ domain corresponds in the figure to rotation about a circle. The
intersection points of the circle with a horizontal bisection line occurs at the antinodes of the
potential $\Phi$. In this figure, at least one particle can always be found at the antinode of
$\Phi$. From the single particle case, we know the antinodes can act as attractors, where individual
particles remain motionless in the $x-\dot{x}$ phase plane. In the case of four particles, shown in
Fig.  \ref{fig:even_odd}(a), two particles may occupy the antinodes of $\Phi$ and the other two
particles oscillate about the nodes of $\Phi$. One might imagine that a rotation of this
configuration by $\pi / 4$ (shown in Fig. \ref{fig:even_odd}(b)) might be a fixed point in the
Poincar\'{e} section of the full phase space, and indeed it is, but it is not a stable configuration
and, unless perfectly configured, it will collapse into the configuration shown in Fig.
\ref{fig:even_odd}(a). For three particles, one particle sits at either antinode and the two
remaining particles compete over the other antinode, as shown in Fig.  \ref{fig:even_odd}(c) and
(d). Which antinode a particle is attracted to depends on the initial conditions.  In both the four
and three particle cases, the particles at the antinodes are stationary. 

Drawing lines between the adjacent average particle positions in the circular topology creates a
regular convex polygon inscribing the circle.  From our description of the particle behavior above,
at least one vertex of the polygon must be at an antinode. For a regular polygon inscribing the
circle with an even number of vertices, each antinode may touch a vertex and it is symmetric under
all rotations obeying this rule. For a polygon with an odd number of vertices, however, only one
vertex can occupy an antinode so that rotations by $\pi/N$ flip the symmetry about a line vertically
bisecting the circle. There are always two unique possible stable configurations when $N$ is odd, but only
one when $N$ is even. This picture changes for even values of $N$ when $N>6$. For $N>6$, the stable
configuration no longer occurs for a pair of particles located at each antinode, but rather for
pairs oscillating on either side of the antinodes much like what is shown in Fig.
\ref{fig:even_odd}(b) but with an extra particle on the top and bottom of the circle. 

\begin{figure*}[htp]
    \centering
    \includegraphics[width=17cm]{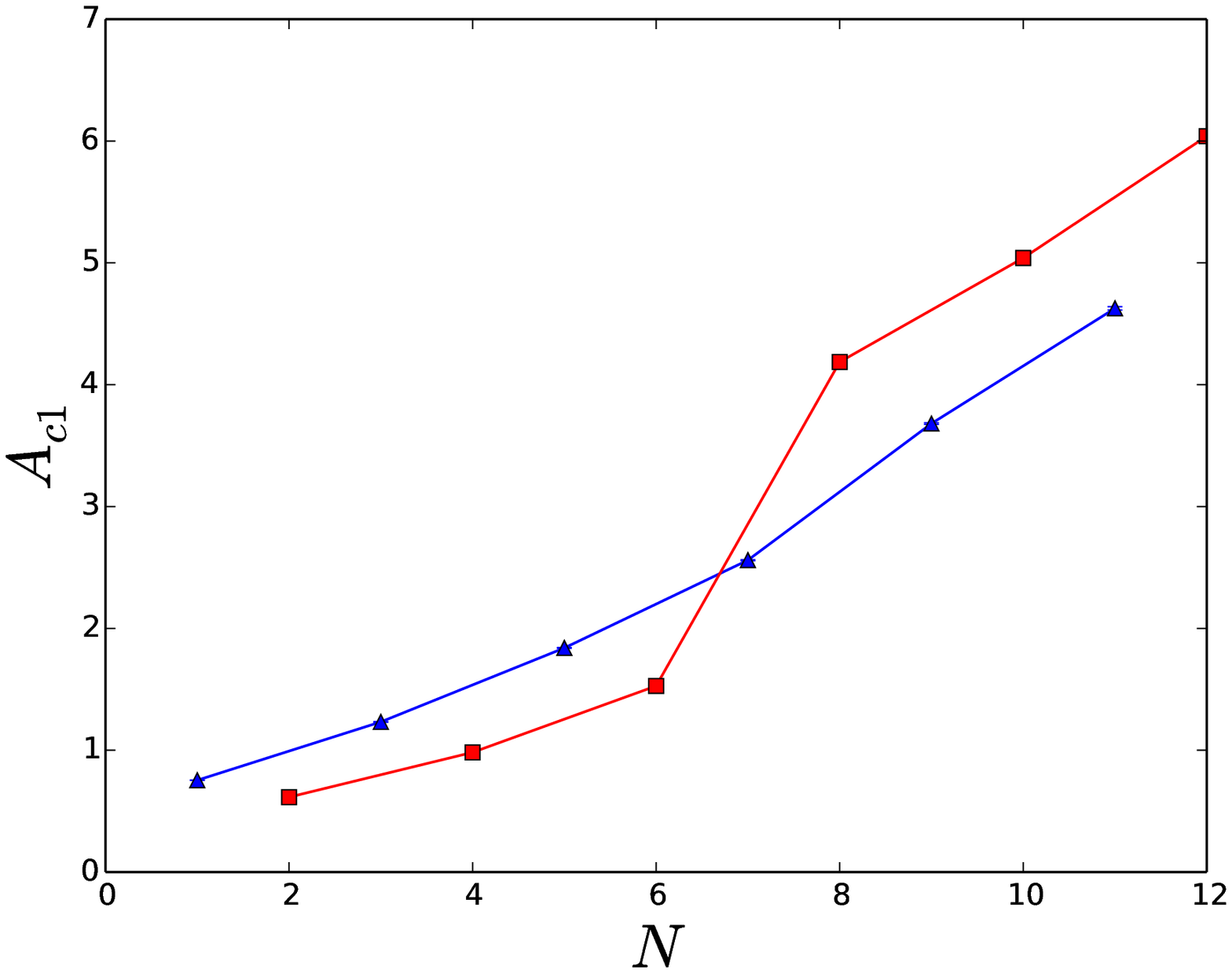}
    \caption{A plot of the value of $A$ at which the system first bifurcates ($A_{c1}$) as a
function of the number of particles in simulation ($N$). Even particle numbers are shown with red
squares and odd particle numbers are shown with blue triangles (color online). A cross-over between
the $N$ odd and $N$ even curves occurs between $N=6$ and $N=7$, which corresponds to the first
bifurcation for the even value of $N$ changing from a Neimark bifurcation to a cyclic
fold bifurcation. The $N$ odd bifurcations are all supercritical flip bifurcations for the values of
$N$ shown here. The upper (lower) error bound is the value of $A$ which we are
certain is after (before) the bifurcation. The error bounds are seen to be quite small in the
Figure.  \label{fig:multi_bif_vals}} 
\end{figure*}

In Fig. \ref{fig:multi_bif_vals}, the value of $A$ at which the first bifurcation occurs is plotted
as a function of $N$, with separate lines for $N$ even (red squares) and $N$ odd (blue triangles). A
striking characteristic of Fig. \ref{fig:multi_bif_vals} is that between $N=6$ and $N=8$, there is a
cross-over point at which the even $N$ line jumps upward and crosses through the odd $N$ line. This
sudden jump in the even $N$ line between $N=6$ and $N=8$ occurs due to a change in type of first
bifurcation. The first bifurcation when $N=6$, as well as the first bifurcations for all lower even
values of $N$, are Neimark bifurcations (a.k.a bifurcation to a torus) in which $N$ stability
multipliers cross the unit circle with nonzero imaginary components. Half of those stability
multipliers ($N/2$) that cross the unit circle have positive imaginary components and the other half
have the complementary negative imaginary components. For $N=8$ and all higher even values of $N$,
the first bifurcation becomes a cyclic fold bifurcation, although the bifurcations for odd values of
$N$ remain supercritical flip bifurcations.

We can qualitatively understand the transition which occurs when $N$ is changed from six to eight in
Fig. \ref{fig:multi_bif_vals} by observing how the bifurcation diagrams, and therefore the stable
attractors, depend on the particle number. In Fig. \ref{fig:multi_bifs}(e), after the first
bifurcation, the particle motions are quenched by the presence of other attractors. For the sake of
discussion we distinguish the two competing sets of attractors based on whether they are found at the far
left or far right edges of the diagrams respectively. Comparing the cases of $N=2$, $N=4$ and $N=6$
in Fig. \ref{fig:multi_bifs} we see similar rightmost attractors that appear abruptly at different
values of $A$ in each case. As $N$ (even) increases, the rightmost attractors increasingly impinge on
the leftmost attractors. This impingement is responsible for the crossing of the lines in Fig.
\ref{fig:multi_bif_vals}.  When $N$ increases from six to eight, the rightmost attractors extinguishes the
left most attractors and becomes the first available attractors for $N\ge 8$ (even). These new first
attractors, previously the rightmost, represents a fundamentally different type of limit cycle in
the full phase space than what was previously first available. Therefore when this attractor first
bifurcates, it falls outside of the original progression, causing the jump in $A_{c1}$ as $N$ is changed
from six to eight shown in Fig. \ref{fig:multi_bif_vals}.

For all of the cases shown in Fig. \ref{fig:multi_bifs}, the system eventually collapses back
into clearly defined attractors which have the form of ``propagating'' trajectories, as was also
observed for the single particle case. These attractors display non-zero net particle motion of one
particle when $N$ is odd, but no net motion when $N$ is even. The particles travel in either the
$\pm x$ direction before a collision-like event. After the ``collision'', they travel in the
opposite direction having exchanged some kinetic energy with the particle with which the collision
occurred. When $N$ is odd the transport of one particle occurs either in the $\pm x$ direction
depending on the initial conditions. There is no possible counter-propagating particle pair for one
of the particles with $N$ odd. 

\begin{figure*}[htp]
    \centering
    \includegraphics[width=17cm]{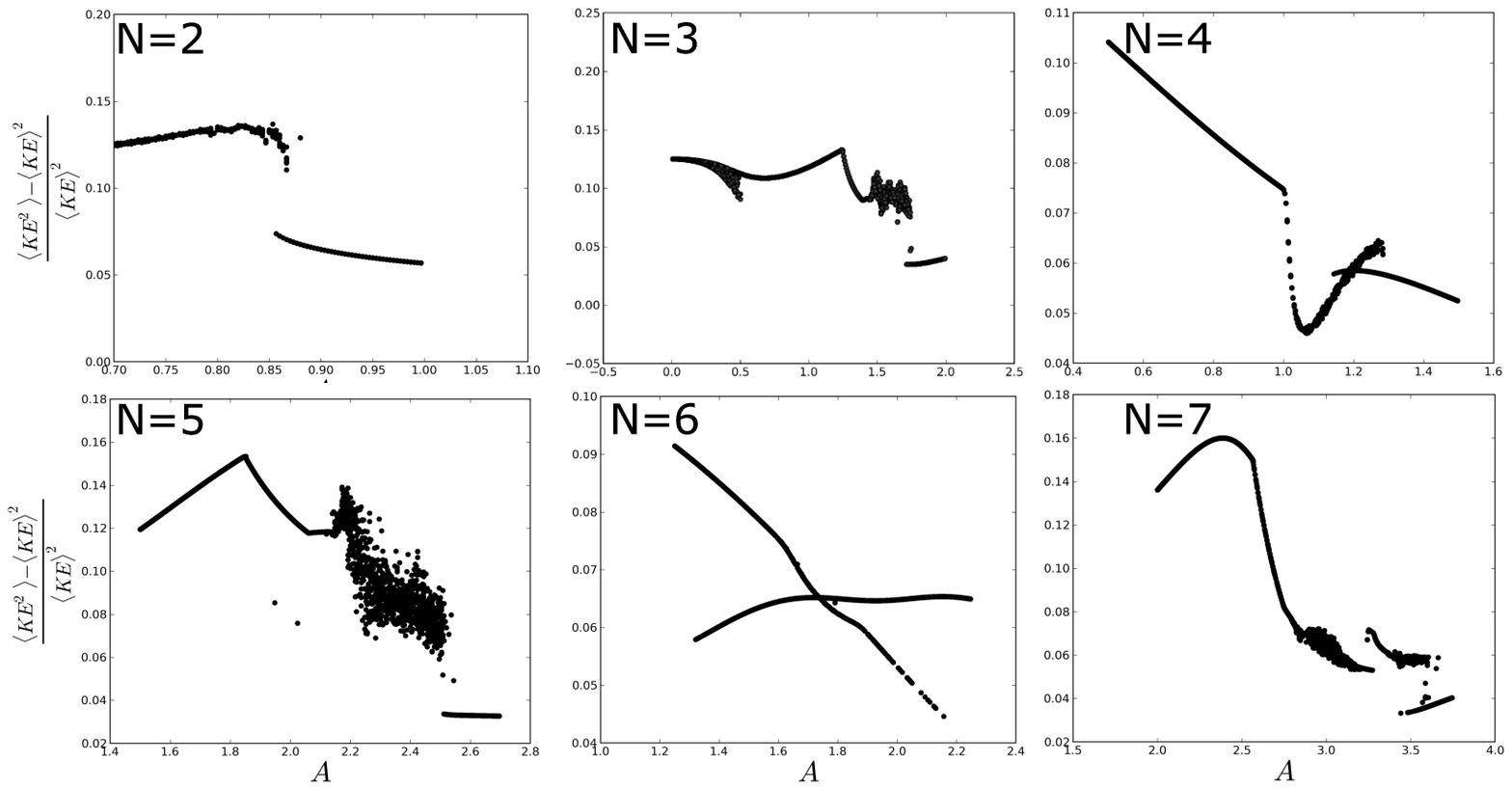}
    \caption{Kinetic energy fluctuations for various particle counts. \label{fig:multi_flucts}}
\end{figure*}

In Fig. \ref{fig:multi_flucts}, the squared fractional deviation of the kinetic energy $\delta_{KE}$
is plotted for all of the bifurcation diagrams shown in Fig. \ref{fig:multi_bifs}. These plots all
exhibit a discontinuity at the point corresponding to transition to a state with the propagating
trajectories. The discontinuity is not as clear in Fig.  \ref{fig:multi_flucts}(e) because (as can be
observed in the corresponding bifurcation diagram Fig. \ref{fig:multi_bifs}(e), the propagating
trajectory begins before the first bifurcation. In Fig. \ref{fig:multi_flucts}(e), the curve starting
below and crossing at $A\approx 1.75$ indicates the values of $\delta_{KE}$ for the propagating
trajectory. 

\subsection{Scaling}
When the periodicity of the system is increased ($n>1$), two distinctly different possibilities
exist. One possibility is an integer concentration,  over a larger periodic system (e.g.  $n=2$ and
$N=4$: $\sigma=2$). The second possibility is a fractional concentration (e.g. $n=2$ and $N=3$:
$\sigma=1.5$). When considering long-range particle interactions, it is reasonable to assume that
increasing the system size might change the dynamics even if the concentration is held fixed. The
first possibility above results in system dynamics very similar to those discussed in the current
paper, whereas the second possibility would result in a completely different symmetry of the system
having very different dynamics.  The effect of period number $n$ on the system dynamics was
examined in the current study by running simulations of equal concentration over larger system sizes
(i.e., large values of $n$). The system dynamics in these larger systems is observed to be
qualitatively the same as for the smaller system sizes reported in the paper, although the exact
values of $A$ for which bifurcations occur is observed to change slightly. We find that for
concentrations larger than two, the effect of scaling the system size while maintaining the
concentration is negligible even when comparing critical $A$ values. 

\section{Conclusion}
We investigate the dynamics of multiple particles with long-range interactions in a STP system by
examining Poincar\'{e} sections and fluctuations of the kinetic energy ($\delta_{KE}$) for different
numbers of particles. Our results are fundamentally interesting because of their importance in
understanding complexity in time-dependent systems. The possible dynamics that exist in a wide range
of different system configurations make the problem challenging, but even in the small area of the
parameter space discussed in this paper we have found a variety of interesting dynamics. For
instance, it is shown that the particle number $N$ influences the stability and the number of
possible final states in a system having integer concentrations. The possible limit cycles of the
system are shown to be sensitive to whether $N$ is even or odd, and the influence of the particle
number on the type of bifurcation is discussed. The squared fractional deviation kinetic energy is
examined as a function of the potential amplitude ($\delta_{KE} (A)$), and it is found to exhibit
interesting features at and near transition points. In particular, discontinuities in
$\frac{d(\delta_{KE})}{dA}$  and $\delta_{KE}$ mark transitions between oscillatory and propagating
modes, respectively. The measure $\delta_{KE}$ may be useful for future experimental investigations
of these systems. 

Our work has demonstrated interesting and complex behaviors of multiple particles with the Coulomb
interaction in a STP potential. In particular, the dynamics of the system is sensitive to particle
concentration and the dynamics can be described by the squared fractional deviation of the particle
kinetic energies.  The latter is particularly valuable for studying bifurcations in real systems.
For example, in the aforementioned studies of the motions of hydrophobic/hydrophilic particles on
the surface of Faraday waves, the particles will interact due to capillary forces caused by their
distortion of the local surface of the water, rather than through the Coulomb interaction, which
leads to particle clustering \cite{faraday_surf_1}. It may be convenient to study this type of
behavior using the squared fractional deviation of the systems kinetic energy because this measure
will decrease in the event of clustering as it measures the effective number of degrees of freedom
\cite{gran_gas}. 

Studies of multiple charged particles in a standing-wave electric curtain and in
acoustic waves are also important areas of research for applications of dust-particle mitigation,
e.g., from a solar panel \cite{liu_jeff}. Charged particles interacting in standing-wave electric
curtains and standing-wave acoustic fields exhibit complicated dynamics that may be illuminated by
studying the squared fractional deviation of the particle kinetic energy.  For example, in
\cite{wu_acoustic_dust} it was observed that for charged particles in a standing-wave acoustic
field, relative motion of smaller particles is faster than that of larger particles, so that the
large particles act as collectors within some agglomeration volume. Any small particles present in
the agglomeration volume are likely to aggregate to a larger particle, and this aggregation is
desirable for applications such as cleaning particles from surfaces. A sweep of the acoustic driving
parameters to find the configuration for which maximal aggregations occurs could clearly be found
and described in terms of the minimal squared fractional deviation of the particle kinetic energies,
again due its measure of th effective number of degrees of freedom. In general, we hope our work may
stimulate further research of STP systems with interacting particles and shed some light on their
complicated and exciting dynamics.

\begin{acknowledgments}
This work was supported by NASA Space Grant Consortium under grant numbers NNX10AK67H, 
NNX08AZ07A, and NNX13AD40A. We also want to acknowledge the Vermont Advanced Computing Core which is supported by
NASA (NNX 06AC88G), at the University of Vermont for providing the high performance computing resources
used for the work in this paper.  
\end{acknowledgments}

\end{document}